\documentclass[11pt]{article}
\usepackage{graphicx}
\usepackage{amssymb,amsmath,amsfonts}

\def\Re{{\cal R \mskip-4mu \lower.1ex \hbox{\it e}\,}}
\def\Im{{\cal I \mskip-5mu \lower.1ex \hbox{\it m}\,}}
\def\ie{{\it i.e.}}

\def\sub#1{_{\lower.25ex\hbox{$\scriptstyle#1$}}}
\def\tev{\,{\ifmmode\mathrm {TeV}\else TeV\fi}}
\def\gev{\,{\ifmmode\mathrm {GeV}\else GeV\fi}}
\def\mev{\,{\ifmmode\mathrm {MeV}\else MeV\fi}}
\def\mpl{\ifmmode M_{pl}\else $M_{pl}$\fi}
\def\mpl{\ifmmode \overline M_{Pl}\else $\bar M_{Pl}$\fi}
\def\to{\rightarrow}

\def\subw{_{\rm w}}
\def\mh{\ifmmode m\sbl H \else $m\sbl H$\fi}
\def\mch{\ifmmode m_{H^\pm} \else $m_{H^\pm}$\fi}
\def\mt{\ifmmode m_t\else $m_t$\fi}
\def\mc{\ifmmode m_c\else $m_c$\fi}
\def\mz{\ifmmode M_Z\else $M_Z$\fi}
\def\mw{\ifmmode M_W\else $M_W$\fi}
\def\mws{\ifmmode M_W^2 \else $M_W^2$\fi}
\def\mhs{\ifmmode m_H^2 \else $m_H^2$\fi}   
\def\mzs{\ifmmode M_Z^2 \else $M_Z^2$\fi}
\def\mts{\ifmmode m_t^2 \else $m_t^2$\fi}
\def\mcs{\ifmmode m_c^2 \else $m_c^2$\fi}
\def\mchs{\ifmmode m_{H^\pm}^2 \else $m_{H^\pm}^2$\fi}
\def\ztwo{\ifmmode Z_2\else $Z_2$\fi}
\def\zone{\ifmmode Z_1\else $Z_1$\fi}
\def\mtwo{\ifmmode M_2\else $M_2$\fi}
\def\mone{\ifmmode M_1\else $M_1$\fi}
\def\tb{\ifmmode \tan\beta \else $\tan\beta$\fi}
\def\xw{\ifmmode x\subw\else $x\subw$\fi}
\def\ch{\ifmmode H^\pm \else $H^\pm$\fi}
\def\lum{\ifmmode {\cal L}\else ${\cal L}$\fi}
\def\inpb{\,{\ifmmode {\mathrm {pb}}^{-1}\else ${\mathrm {pb}}^{-1}$\fi}}
\def\infb{\,{\ifmmode {\mathrm {fb}}^{-1}\else ${\mathrm {fb}}^{-1}$\fi}}
\def\epem{\ifmmode e^+e^-\else $e^+e^-$\fi}
\def\ppb{\ifmmode \bar pp\else $\bar pp$\fi}
\def\bsg{\ifmmode B\to X_s\gamma\else $B\to X_s\gamma$\fi}
\def\bsll{\ifmmode B\to X_s\ell^+\ell^-\else $B\to X_s\ell^+\ell^-$\fi}
\def\bstt{\ifmmode B\to X_s\tau^+\tau^-\else $B\to X_s\tau^+\tau^-$\fi}
\def\lamt{\ifmmode \tilde\lambda\else $\tilde\lambda$\fi}
\def\shat{\ifmmode \hat s\else $\hat s$\fi}
\def\that{\ifmmode \hat t\else $\hat t$\fi}
\def\uhat{\ifmmode \hat u\else $\hat u$\fi}

\newskip\zatskip \zatskip=0pt plus0pt minus0pt
\def\matth{\mathsurround=0pt}
\def\lsim{\mathrel{\mathpalette\atversim<}}

\def\atversim#1#2{\lower0.7ex\vbox{\baselineskip\zatskip\lineskip\zatskip
  \lineskiplimit 0pt\ialign{$\matth#1\hfil##\hfil$\crcr#2\crcr\sim\crcr}}}

\def\grtsim{\,\,\rlap{\raise 3pt\hbox{$>$}}{\lower 3pt\hbox{$\sim$}}\,\,}
\def\lsim{\,\,\rlap{\raise 3pt\hbox{$<$}}{\lower 3pt\hbox{$\sim$}}\,\,}

\renewcommand{\thefootnote}{\fnsymbol{footnote}}

\begin{document} \begin{titlepage}
\rightline{\vbox{\halign{&#\hfil\cr
%&DRAFT\cr
&SLAC-PUB-13222\cr
%&March 2006\cr
}}}
\begin{center}
\thispagestyle{empty} \flushbottom { {
\Large\bf Identification of the Origin of Monojet Signatures at the LHC  
\footnote{Work supported in part
by the Department of Energy, Contract DE-AC02-76SF00515}
\footnote{e-mail:
rizzo@slac.stanford.edu}}}
\medskip
\end{center}

\centerline{Thomas G. Rizzo}
\vspace{8pt} 
\centerline{\it Stanford Linear
Accelerator Center, 2575 Sand Hill Rd., Menlo Park, CA, 94025}

\vspace*{0.3cm}

\begin{abstract}
Several new physics scenarios can lead to monojet signatures at the LHC. If such events are observed above the 
Standard Model background it will be important to identify their origin. In this paper we compare and contrast 
these signatures as produced in two very different pictures: vector or scalar unparticle production in the 
scale-invariant/conformal regime and graviton emission in the Arkani-Hamed, Dimopoulos and Dvali extra-dimensional 
model. We demonstrate that these two scenarios can be distinguished at the LHC for a reasonable range of model 
parameters through the shape of their respective monojet and/or missing $E_T$ distributions. 
\end{abstract}

%\vskip0.45in
%\begin{center}

%\end{center}

\renewcommand{\thefootnote}{\arabic{footnote}} \end{titlepage} 

%
%
%
%%%%%%%%%%%%%%%%%%%%%%%%%%%%%%%---- Put text here

\section{Introduction and Background}

The LHC will turn on during 2008 and it is generally expected that new physics beyond the Standard Model(SM) will be discovered at some 
point thereafter once significant luminosity is accumulated and the the two detectors are sufficiently well-understood. What new physics 
signatures will be observed and how will they 
be grounded within a specific theoretical framework? Clearly once the new physics is found, our primary goal will  
be to uncover its origin and to identify the underlying model which generates it. This can sometimes be confusing as in many cases 
several kinds of new physics can lead to quite similar signatures at colliders. Supersymmetry{\cite {SUSY}} and Universal Extra Dimensions{\cite {UED}} 
provide us with one well-known example of this possible model confusion which has been much discussed in the recent literature{\cite {confusion}}.  
The ability of the LHC to differentiate models with similar signatures may become the the most important issue once new physics is discovered. 

In this paper we will consider another example of this kind of potential confusion which may arise at the LHC between two very different 
scenarios: graviton emission in the extra-dimensional model 
of Arkani-Hamed, Dimopoulos and Dvali(ADD){\cite {ADD}} and the production of scalar/vector unparticles in the model of Georgi{\cite {georgi}} 
in the scale-invariant/conformal regime{\cite {conformal}} where the unparticle can be treated as both `massless' and stable. Here, by a `massless' 
unparticle we will mean one that has a null threshold mass parameter, \ie, $\mu^2=0$, which determines the {\it minimum} allowed value of its possible 
squared 4-momentum{\cite {conformal}}, \ie, $\mu^2 \leq P^2$. (Recall that for unparticles the value of $P^2$ actually takes on a continuous range 
which is integrated over to obtain observable cross sections.) 
Both of these new physics models can lead to a monojet signal, \ie, a single jet plus missing $E_T$ (MET) with balancing transverse momenta,  
at the LHC. In the ADD case, this 
scenario has been studied in some detail in the classic work by Vacavant and Hinchliffe{\cite {hinch}} within the ATLAS{\cite {ATLAS}} setting, which we 
will use as a guide for the present analysis.{\footnote {CMS{\cite {CMS}} has also performed a more recent but comparable analysis in the single photon channel.}} 
Assuming that monojet signals above the SM backgrounds are indeed observed at the LHC we will demonstrate that the 
shapes of the corresponding excess jet and/or missing $E_T$ distributions are qualitatively quite distinctive in these two cases. This will  
allow these two classes of models to be distinguished at the LHC for a range of parameters provided that sufficient integrated luminosity is available. 

As emphasized by Vacavant and Hinchliffe{\cite {hinch}, observing an excess in the monojet channel relies on our thorough understanding of the SM 
background. At low $E_T$, this background can be dominated by QCD/jet energy mis-measurements. However, at high $E_T$ it is largely dominated by $Z+j$ 
production followed by the decay $Z\to \nu \bar \nu$. Fortunately, in practice, it appears that neither ATLAS nor CMS{\cite {both}} will need to rely 
solely on Monte Carlo estimates to fully understand this MET plus jet(s) background.  At high $E_T$  one can instead employ the `standard candle' approach 
where one makes use of the same production channel but now with the $Z$ decaying leptonically, \ie, $Z\to e^+e^-/\mu^+\mu^-$. These easily 
measured rates can then be corrected for differences in the branching fractions as well as for acceptances and efficiencies to obtain the $Z$-induced SM monojet 
background{\cite {tevatron}} directly from data. Other sources of background, such as $Z+V$ production with $Z\to \nu \bar \nu,~V\to jj$ with the two jets 
coalescing due to the boost, can also be handled in a similar fashion though we expect them to be about an order of magnitude smaller in rate. 
In the analysis presented below we will assume that this procedure works so that these  
backgrounds will eventually be well understood in the very high $E_T$ region above 500 GeV, corresponding to an $M_{eff}=E_{T_j}+\rm{MET}\geq 1$ TeV upon which 
we will concentrate. 
We will not include in the present analysis possible additional information that may be obtainable from the lower $E_T$ region where the jet energy 
mis-measurements can be a potentially large source of SM backgrounds. The resulting SM background estimates that we obtain and will employ below are 
found to be essentially the same as, though perhaps $10-20\%$ larger than, those found by Vacavant and Hinchliffe{\cite {hinch}}. 
Of course the background in this analysis is not the issue at hand. Our goal is to inquire whether or not the ADD and unparticle predictions can be distinguished 
provided a well-defined signal over background is already clearly observed.

Our procedure is rather straightforward: First, we will demonstrate that the overall {\it shape} of the normalized monojet $E_T$ distribution predicted by 
the ADD model is essentially independent of the number of extra dimensions in the relevant parameter range. Next we will show that the predictions for the 
corresponding normalized monojet $E_T$ spectra produced by either scalar and vector unparticles lie in a rather narrow band. By comparing these two sets 
of distributions it will then become clear that for sufficiently high integrated luminosities, $\sim 100 fb^{-1}$, these two predictions will be easily 
isolated from one another in $E_T$ space so that we can distinguish these two classes of models at the LHC by using the data collected above $E_T=500$ GeV.

\section{Analysis}

Let us begin by considering the monojet signal{\cite {GRW}} in the ADD model which arises from graviton Kaluza-Klein tower ($G$) emission in 
the following processes: $q\bar q \to gG,~gg\to gG$ and $q(\bar q)g\to q(\bar q)G$. The expressions for these parton-level cross sections 
are given in full detail in Ref.{\cite {GRW}}. In the original ADD scenario, the resulting cross section expressions depend upon only 2 parameters: 
the number of extra dimensions, $2\leq \delta \leq 7$, and the value of the $D=4+\delta$-dimensional Planck scale, $M_D$; these subprocess level cross 
sections are observed to scale as $\sim M_D^{-(2+\delta)}$. Since the ADD model is only an effective theory below a cut off scale $\Lambda \sim M_D$, 
in performing 
the integrations necessary to obtain the relevant LHC cross sections it is unclear how to treat the kinematic region where the partonic center of mass 
energy, $\hat s$, is in excess of $\Lambda^2$ {\it unless} one has a more fundamental theory to rely upon. There are two common ways that have been used to 
address this issue discussed in the literature{\cite {hinch,GRW}}. First, we could simply ignore this problem and perform the necessary integration; a second 
possibility is to make a hard truncation in the integration at $\hat s =\Lambda^2$, both of which are unrealistic. The former choice will, of course, lead to 
a larger event rate 
but one finds that these two approaches yield qualitatively similar but quantitatively different numerical results.  

In performing 
our analysis, we will instead follow a better approach and make use of a {\it realistic} underlying fundamental theory based on non-perturbative General 
Relativity in extra dimensions.  This approach leads directly to non-Gaussian fixed-points in the gravitational coupling that predicts the existence of 
a gravitational form factor{\cite {us}} which naturally controls the large $\hat s$ 
part of the cross section, extending the range of validity of the partonic cross section above $\hat s \sim M_D^2$ and maintains unitarity at high energies. It is 
important to note that this is not an ad-hoc use of an {\it arbitrary} form factor as the underlying dynamics dictates the form factor structure and power 
behavior as well as the form 
factor scale itself up to a factor close to unity. For 
numerical purposes we will use a fixed form factor scale of $\Lambda_{FF}=8$ TeV in this analysis. One finds, however, that varying this particular 
choice will make little numerical difference in the qualitative nature of the results presented below, and as an explicit examination reveals, since the form 
factor only leads to 
modifications in the extreme high end of the various $E_T$ distributions where there are very limited statistics as we will see. 

\begin{figure}[htbp]
\centerline{
\includegraphics[width=7.5cm,angle=90]{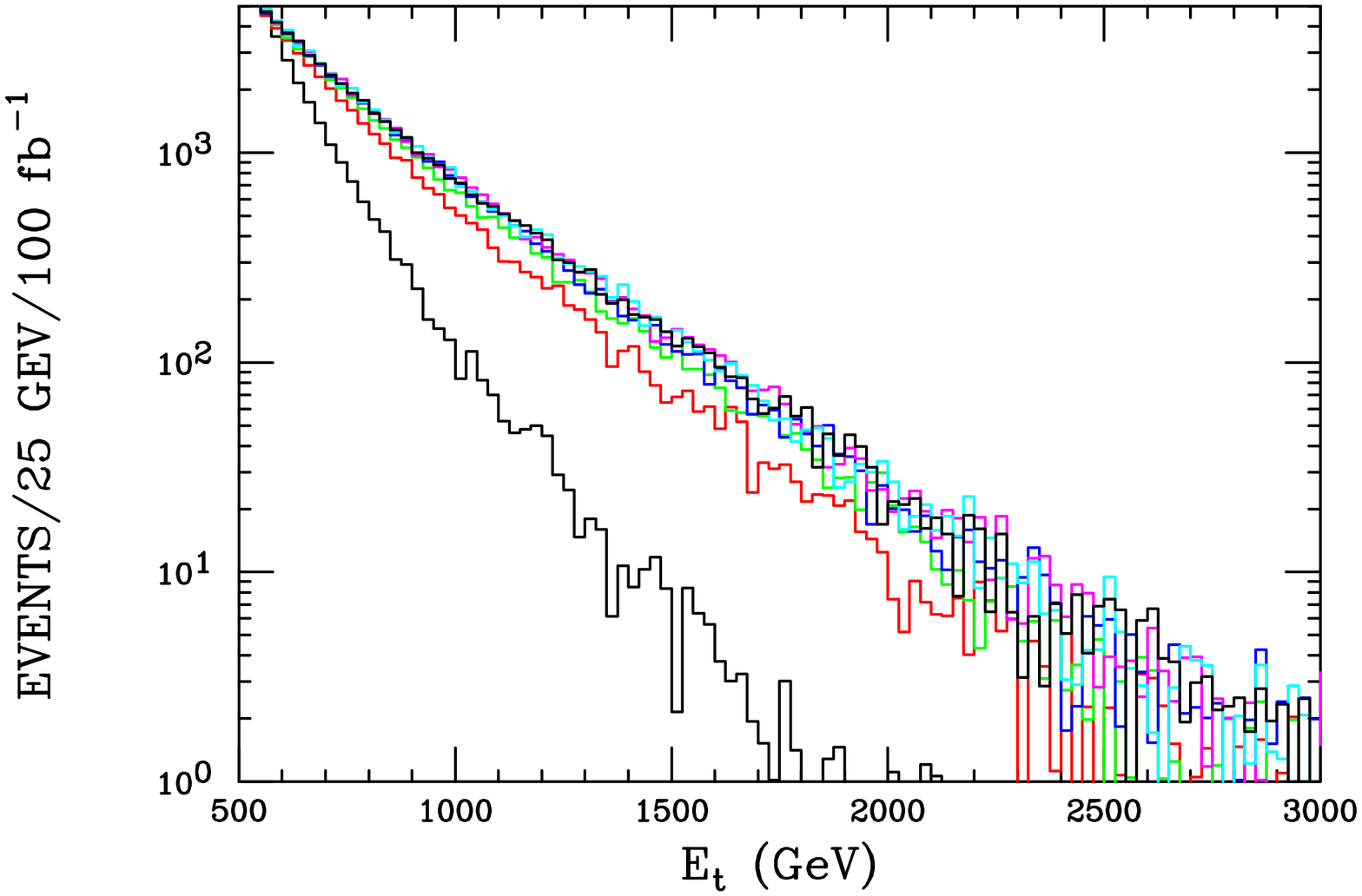}}
\vspace*{0.1cm}
\centerline{
\includegraphics[width=7.5cm,angle=90]{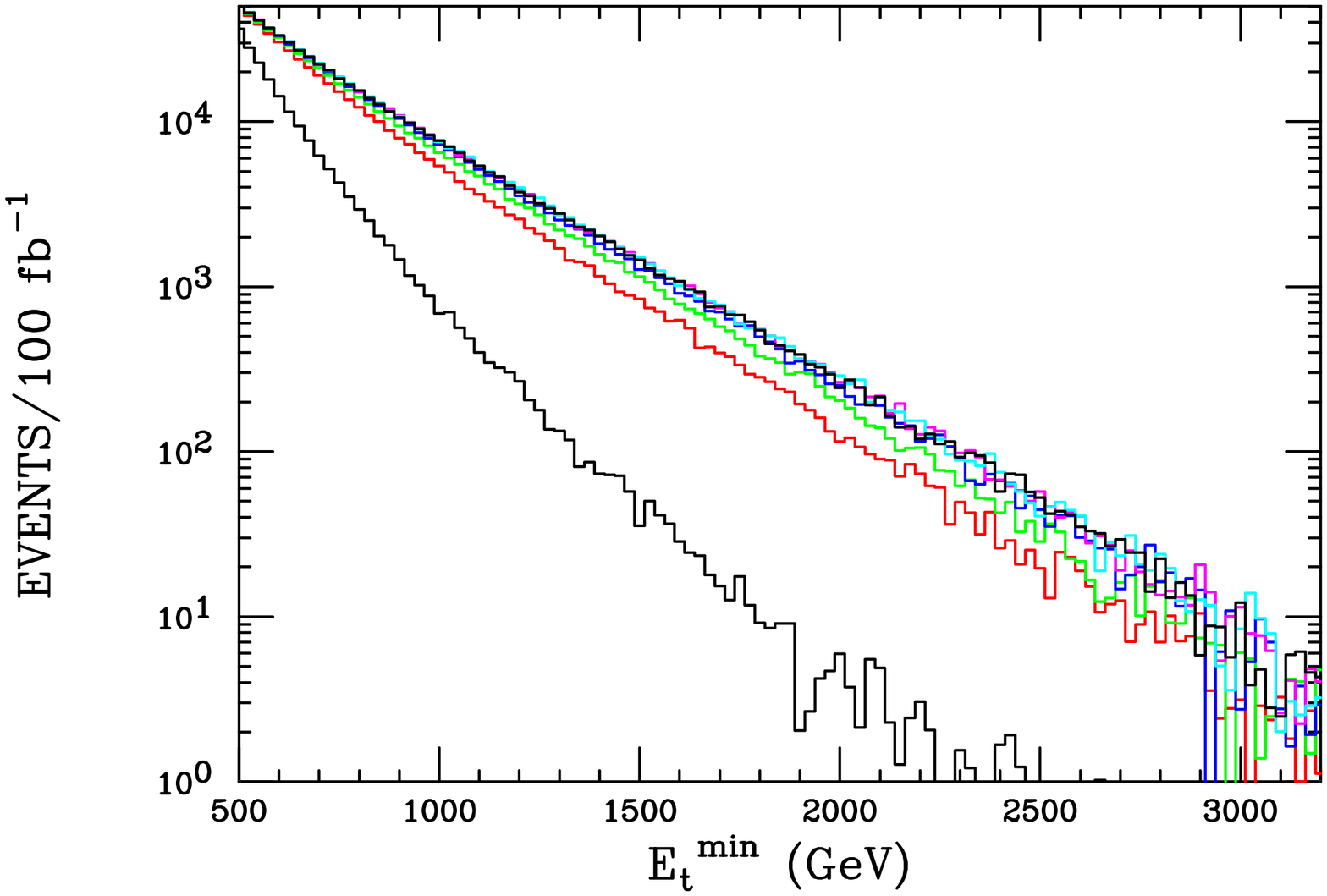}}
\vspace*{0.1cm}
\caption{Event rate for the monojet signal induced by graviton emission in the ADD model assuming $2\leq \delta \leq 7$, in steps of unity from bottom to top, 
corresponding to the red, green, blue, magenta, cyan and black histograms,  
as a function of the jet $E_T$ in the top panel or above a minimum jet $E_T$ cut in the lower panel. Note that a cut on the jet rapidity of 
$|\eta_j|\leq 3$ has been applied in all cases. As discussed in the text the values of $M_D$ have been adjusted in each case so as to give the 
same prediction as that for $M_D=4$ TeV with $\delta=2$ at an $E_T$ value 
of 500 GeV. Note that for comparison purposes these ADD predictions are for pure signal {\it only}. The SM background is represented in either panel 
by the lowest black histogram.}
\label{fig1}
\end{figure}
\begin{figure}[htbp]
\centerline{
\includegraphics[width=7.5cm,angle=90]{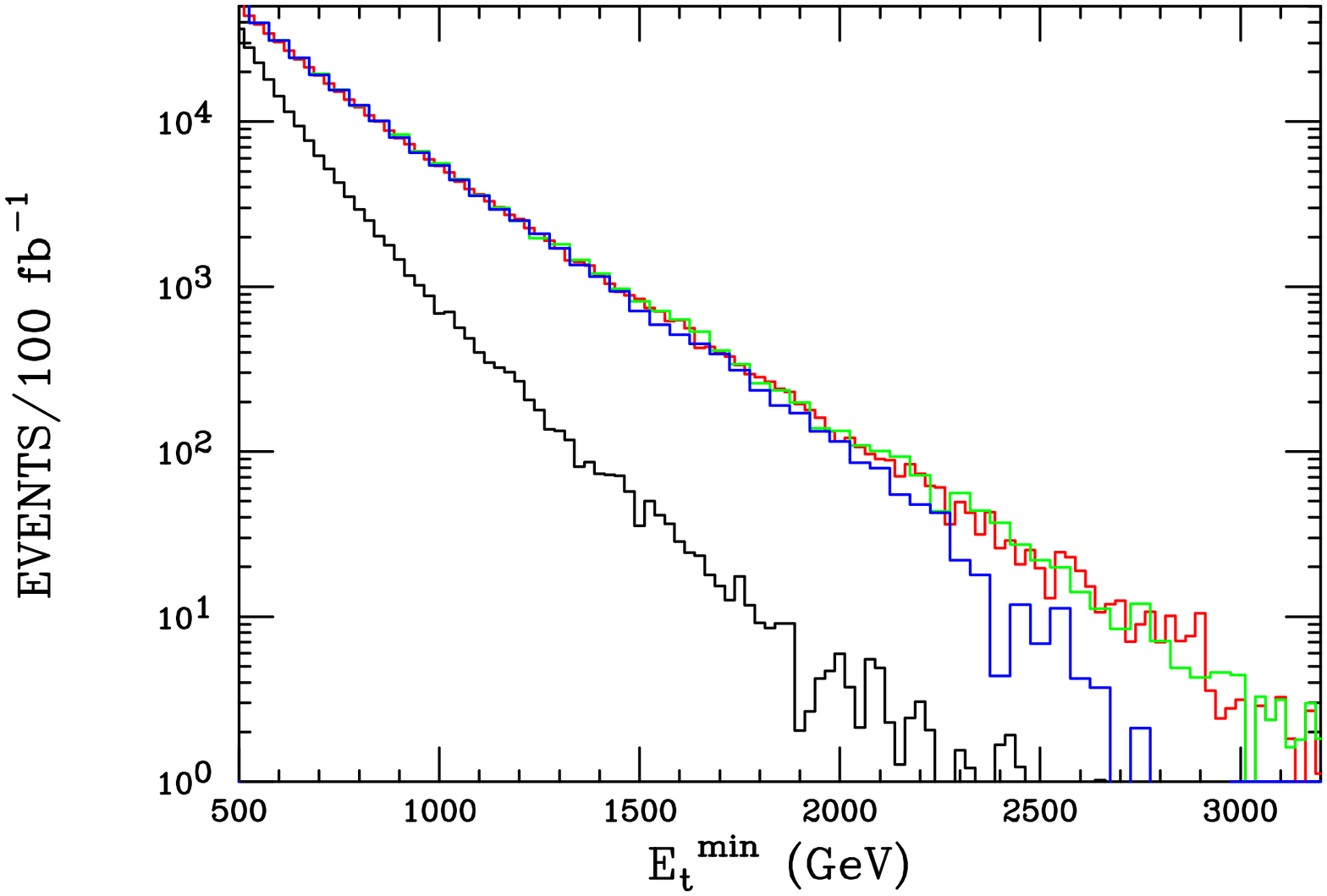}}
\vspace*{0.1cm}
\caption{Event rate for the monojet signal induced by graviton emission in the ADD model assuming $\delta=2$ above a minimum jet $E_T$ as in the previous 
figure along with the SM background. 
Here we compare the results obtained using the form factor prescription which we follow here (red), to that of integration over the entire $\hat s$ range (green) 
and to that where a strong cut off of $\Lambda=1.5M_D$ has been applied (blue). 
Note that as in the previous figure a cut on the jet rapidity of $|\eta_j|\leq 3$ has been applied in all cases.
}
\label{fig0}
\end{figure}

Since our goal is to differentiate monojet signals arising in different models by the {\it shapes} of their jet $E_T$ distributions, it 
is instructive to first compare the ADD model `with itself', \ie, to compare the ADD predictions for different values of $\delta$ while simultaneously 
varying the value of $M_D$ (as a function of $\delta$) so that the models all predict the same cross section at some fixed value of the jet $E_T$. 
This allows us to directly compare the predicted shape for the signal monojet spectrum for each value of $\delta$ without having to worry about the 
overall normalization of the spectrum. Note that in all cases the subprocess cross sections for graviton emission are seen to 
grow rapidly with $\hat s$ and so can naturally lead to large event excesses at high $E_T$. As an example, let us assume a reference value for the 
monojet cross section corresponding to the choice of $\delta=2$ 
with $M_D=4$ TeV. For all larger values of $\delta$ we will then adjust the associated value of $M_D$ such that identical cross sections are obtained 
when evaluated at a jet energy of $E_T=500$ GeV. Note that there is nothing particularly special about this choice of $E_T$ value other than it allows us 
to have a reasonable signal-to-background ratio for all values of $\delta$ for the integrated luminosities that we will assume in our analysis. Other 
choices for this cross-over point in the ADD case would yield qualitatively similar results since we are only probing the various shapes of the $E_T$ 
distributions here. We next determine the resulting LHC monojet event rates for each case as a function of jet $E_T$  while simultaneously demanding that 
the monojet be more or less central in rapidity, \ie, $|\eta_j|\leq 3$, as assumed by Vacavant and Hinchliffe{\cite {hinch}}; to be specific, we employ 
the CTEQ6.6M PDFs{\cite {cteq}} throughout and will assume a fixed integrated luminosity of 100 $fb^{-1}$. In a similar manner, we can also ask for the 
cumulative number of events {\it above} a minimum cut on the jet $E_T$ which probes a somewhat different aspect of the jet $E_T$ spectrum. As we will see, 
the additional statistics in this distribution will allow for an improved separation of models.

\begin{figure}[htbp]
\centerline{
\includegraphics[width=7.5cm,angle=90]{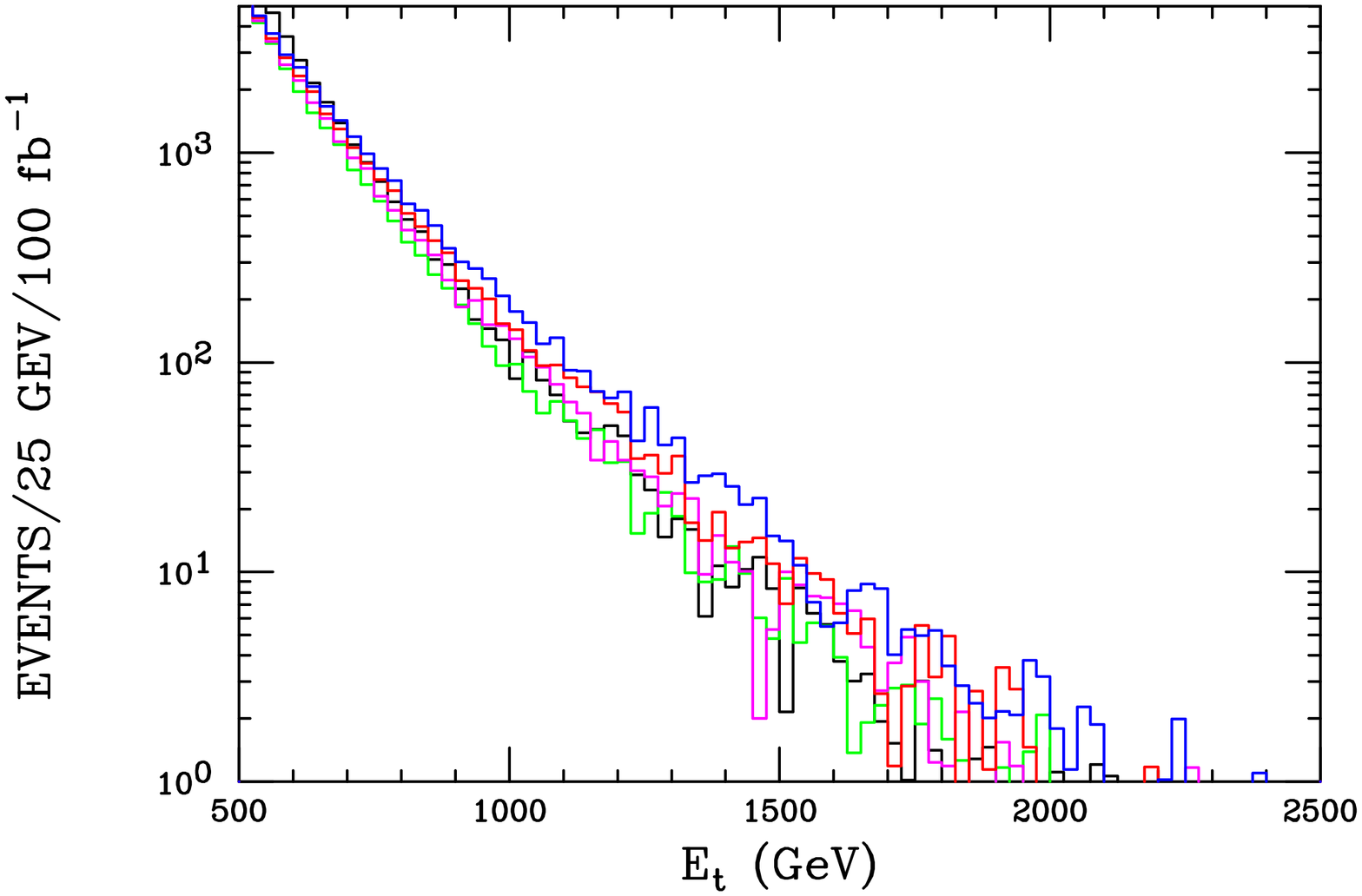}}
\vspace*{0.1cm}
\centerline{
\includegraphics[width=7.5cm,angle=90]{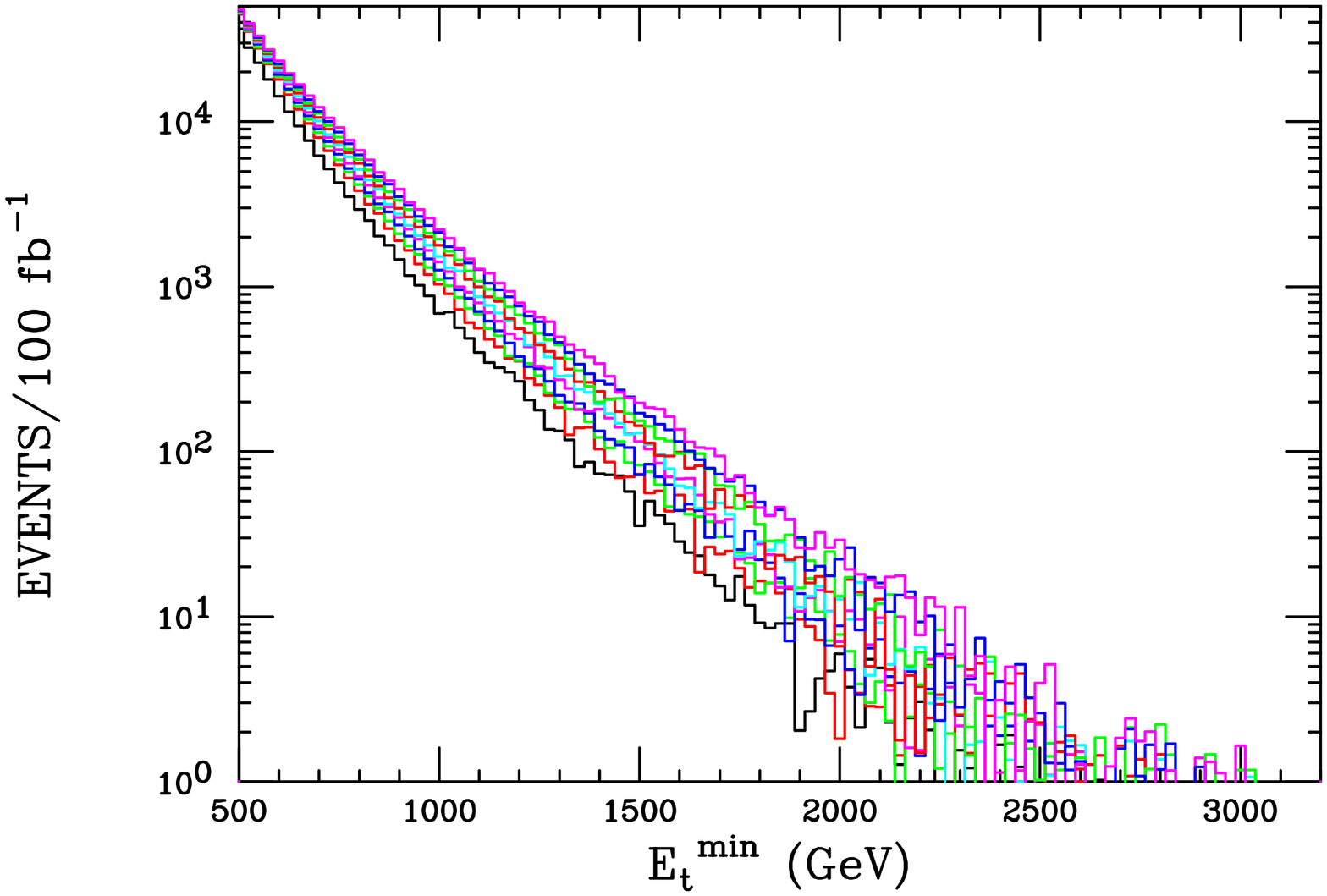}}
\vspace*{0.1cm}
\caption{Same as Fig.1, but now for vector unparticles with, from bottom to top on the left-hand side of each panel, 
$d=1.1$ to $d=1.9$ in steps of 0.1(0.2) in the lower(upper) panel.}
\label{fig2}
\end{figure}

The results 
of this analysis together with an estimate of the SM $Z+j$ induced backgrounds are shown in Fig.~\ref{fig1} for $2\leq \delta \leq 7$ assuming jet 
$E_T$'s larger than 500 GeV. Here we see that while the ADD model for the chosen set of parameter ranges  
clearly produces a significant signal above the SM background it is very difficult to determine the value of $\delta$ itself since all of the 
predicted distributions lie rather close to one another{\footnote {The results for $\delta=2$ are seen to lie slightly below the others so that this case may be 
confidently distinguished from the others at somewhat higher integrated luminosities. It is important to note, however, that the predictions for larger 
values of $\delta$ lead to slightly harder distributions.}}. This is seen to be an even stronger conclusion for the case of the $E_T^{cut}$ distributions.
This reproduces the well-known result{\cite {hinch}} 
that at the LHC it is essentially impossible, or at least extremely difficult, to determine both the values of $M_D$ and $\delta$ simultaneously from 
the observation of a monojet excess at a {\it fixed} value of $\sqrt s$.  However, for our purposes, this is a {\it positive} result since our goal is to 
distinguish the set of {\it all} ADD predictions from those of other models, specifically, those arising from unparticles. The ADD prediction for the shape 
of these $E_T$ and $E_T^{cut}$ distributions are 
thus found not to be very sensitive to the specific value of $\delta$, especially when $3 \leq \delta$. Note that if the value of 
$E_T=500$ GeV, chosen as the cross-over point where the various ADD predictions yield the same cross section, was taken to be somewhat larger then any 
determination of the value of $\delta$ would be only be made more difficult since the divergences in the observed spectra would be reduced. Lowering 
the cross-over point would increase the lever arm somewhat but, as we can easily see from this figure, would at most only be useful in separating the 
$\delta=2$ case from the other choices for $\delta$. 

\begin{figure}[htbp]
\centerline{
\includegraphics[width=7.5cm,angle=90]{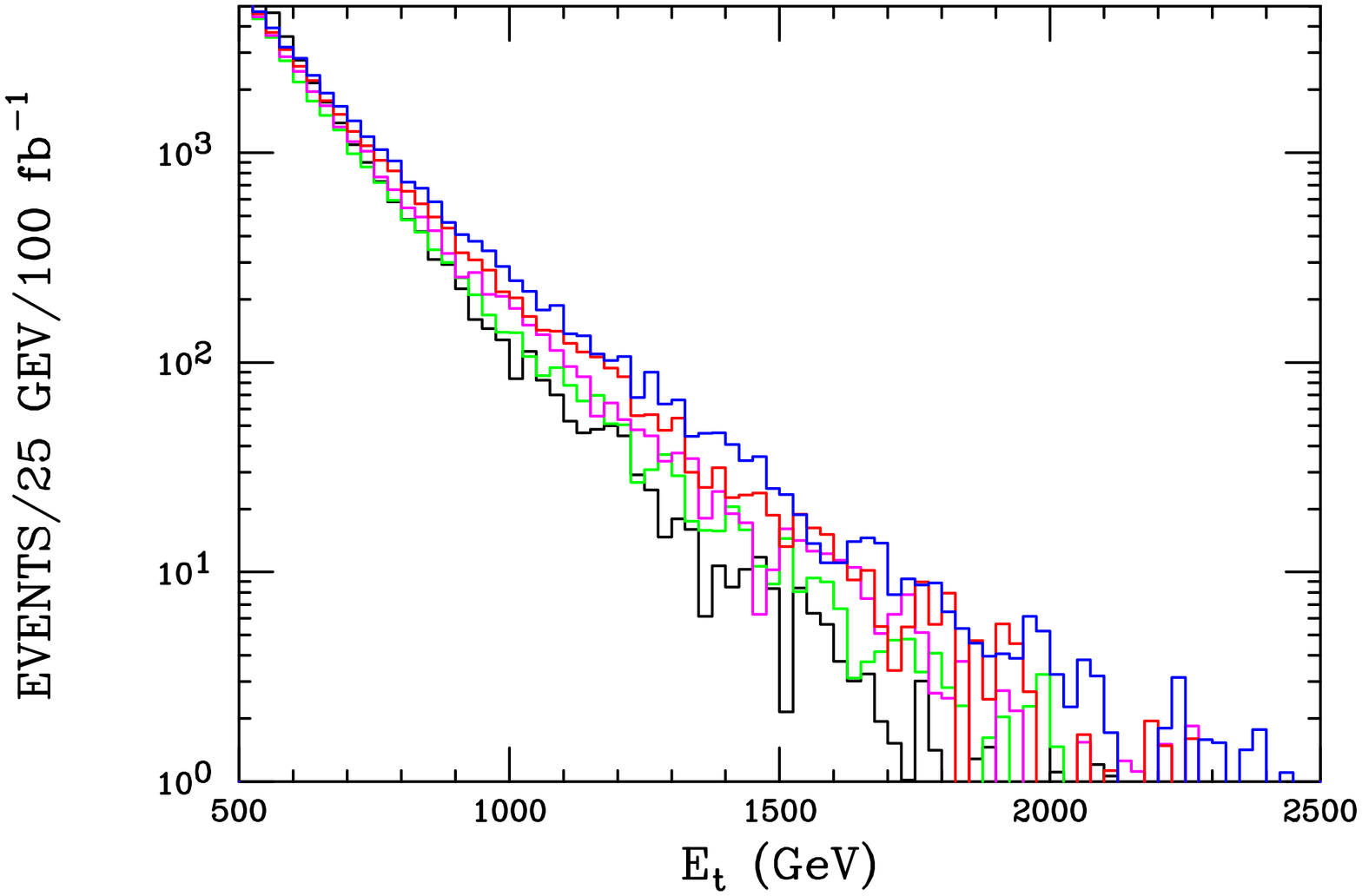}}
\vspace*{0.1cm}
\centerline{
\includegraphics[width=7.5cm,angle=90]{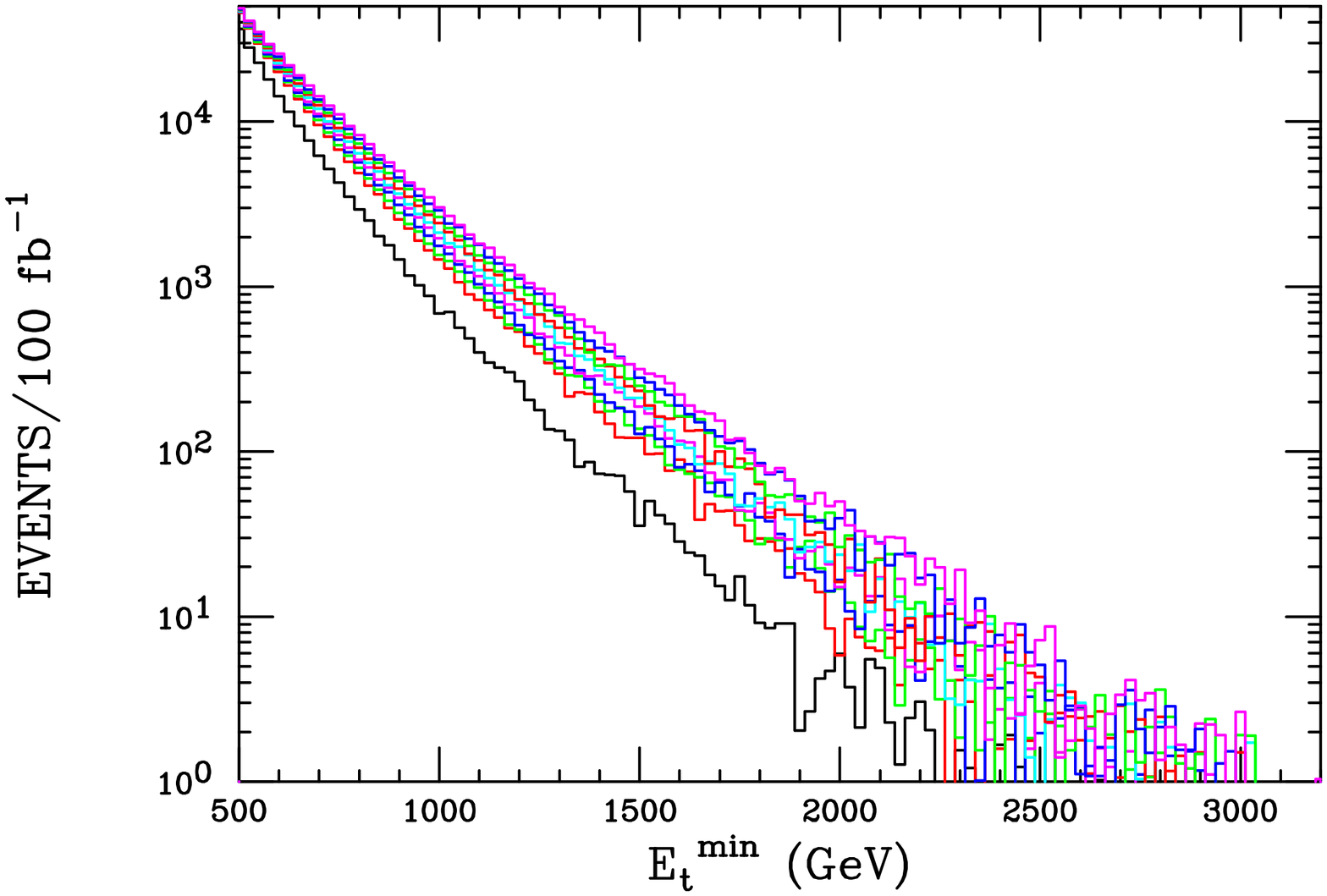}}
\vspace*{0.1cm}
\caption{Same as Fig.1, but now for $r=0$ scalar unparticles with, from bottom to top, $d=1.1$ to $d=1.9$ in steps of 0.1(0.2) in the lower(upper) panel.}
\label{fig3}
\end{figure}

In order the compare the standard approaches to the large $\hat s$ problem in ADD and what we obtain here, a direct comparison for the case of $\delta=2$ is made 
as described in Fig.~\ref{fig0}. Here we see that the predictions of form-factor analysis lies fairly close to those obtained by integration over the full 
kinematic range while those with a hard cut off are somewhat softer at the very high/low statistics end as would be expected. In all cases the shape/slope of this 
distribution is seen to be almost identical for an jet $E_T$ cut below $\sim 2.3$ TeV.

When considering the predictions of the unparticle model, with which we wish to compare the ADD scenario, the parameter space is somewhat more varied and 
there are many possibilities. Here, to be specific, we will consider the case of `massless' unparticles, as described above, which are assumed to be  
either scalar (spin-0), $U_S$, or vector (spin-1), $U_V$, and in either case 
are sufficiently stable so that they can lead to a collider missing $E_T$ signature. However, we have checked that by adding a small 
unparticle threshold mass,  $\mu \lsim 100$ GeV, so that $P^2 \geq \mu^2$, we will not significantly alter the numerical results presented below as long 
as the relative stability of the unparticle on collider scales is also maintained. This is clear since for these $2\to 2$ processes small threshold masses 
for the unparticles will only result in a small reduction in the size of the relevant phase space which is of order $\sim \mu^2/\shat$. 
Having made these particular choices, several other parameter 
options still can remain. If we take the simplest possibility and assume that the unparticle couples as 
a SM gauge singlet to only one operator constructed of only single type of SM field, then the resulting interaction will depend upon only an overall scale, 
$\Lambda$, as well as the value of the effective unparticle anomalous dimension, $d$, which is restricted to lie in the range $1<d<2$. If, however, 
the unparticle is allowed to couple to several different SM operators, then the relative strengths of these interactions, 
$r_i$, can also become quite relevant. Here, to be specific, we will assume that the spin-0 unparticles couple to the relevant SM fields via  
\begin{equation}
{1\over {\Lambda^{d-1}}}\sum_q \bar q q U_S +r {1\over {\Lambda^d}} G_{\mu\nu}^a G_a^{\mu\nu}U_S\,,
\end{equation}
for all quark fields $q$ and where $G_{\mu\nu}^a$ is the gluon field strength tensor; we will consider the two cases $r=0,1$ below as representative 
examples. We will further assume that the spin-1 unparticles couple only to the SM quarks via 
\begin{equation}
{1\over {\Lambda^{d-1}}}\sum_q \bar q \gamma_\mu q U_V^\mu\,,
\end{equation}
also in a universal manner. The parton-level cross sections for the relevant processes $q\bar q \to gU_V$, $q(\bar q)g \to q(\bar q)U_V$ and $gg\to gU_S$ 
can be found in Ref.{\cite {keung}}. Correspondingly, we find that the (reduced) spin and color-averaged squared matrix elements for the remaining processes 
of interest are given, in a slightly modified version of the notation of Ref.{\cite {keung}}, by 
\begin{eqnarray}
 |\bar M|^2(q\bar q \to gU_S) &=& {{16\pi \alpha_s}\over {9\hat u \hat t \Lambda^2}} (\hat s^2-(P^2)^2)   \\ \nonumber
 |\bar M|^2(qg \to qU_S) &=&  {{2\pi \alpha_s}\over {-3\hat u \hat s \Lambda^2}} (\hat t^2-(P^2)^2)  \,,
\end{eqnarray}
where $P^2$ is the unparticle invariant mass, as defined above, which is integrated over to obtain a final cross section result. 
A similar expression is found to hold in the case of the $\bar qg$ initial state. The corresponding differential cross sections are then given by  
\begin{equation}
{{d^2\sigma}\over{d\hat t dP^2}}={{1}\over {16\pi \hat s^2}} {{A_d}\over {2\pi}}|\bar M|^2 \Big({{P^2}\over {\Lambda^2}}\Big)^{d-2}\,,
\end{equation}
where $\hat t$ is the usual subprocess-level Mandelstam variable and $A_d$ is the familiar unparticle numerical phase space factor as given by 
Georgi{\cite {georgi}}. 
It is interesting to note that the scaling behavior of the unparticle cross sections for both the unparticle vectors and $r=0,1$ scalars is 
such that they will lead to essentially the same $E_T$ dependence as the SM background in the limit when $d\to 1$. This is not surprising since in 
that limit the unparticles behave in a manner similar to the more conventional SM boson fields.  
In keeping with the ADD distribution shape analysis above, we will in all cases adjust the value of the scale $\Lambda$ so that the unparticle monojet 
cross section at $E_T=500$ GeV is the same as that for the ADD model with $M_D=4$ TeV with $\delta=2$ at the same $E_T$. In order to do this it is 
obvious that different values of $\Lambda$ will need to be chosen as both $d$ and $r$ are varied for both the scalar and vector unparticle cases 
independently.

\begin{figure}[htbp]
\centerline{
\includegraphics[width=7.5cm,angle=90]{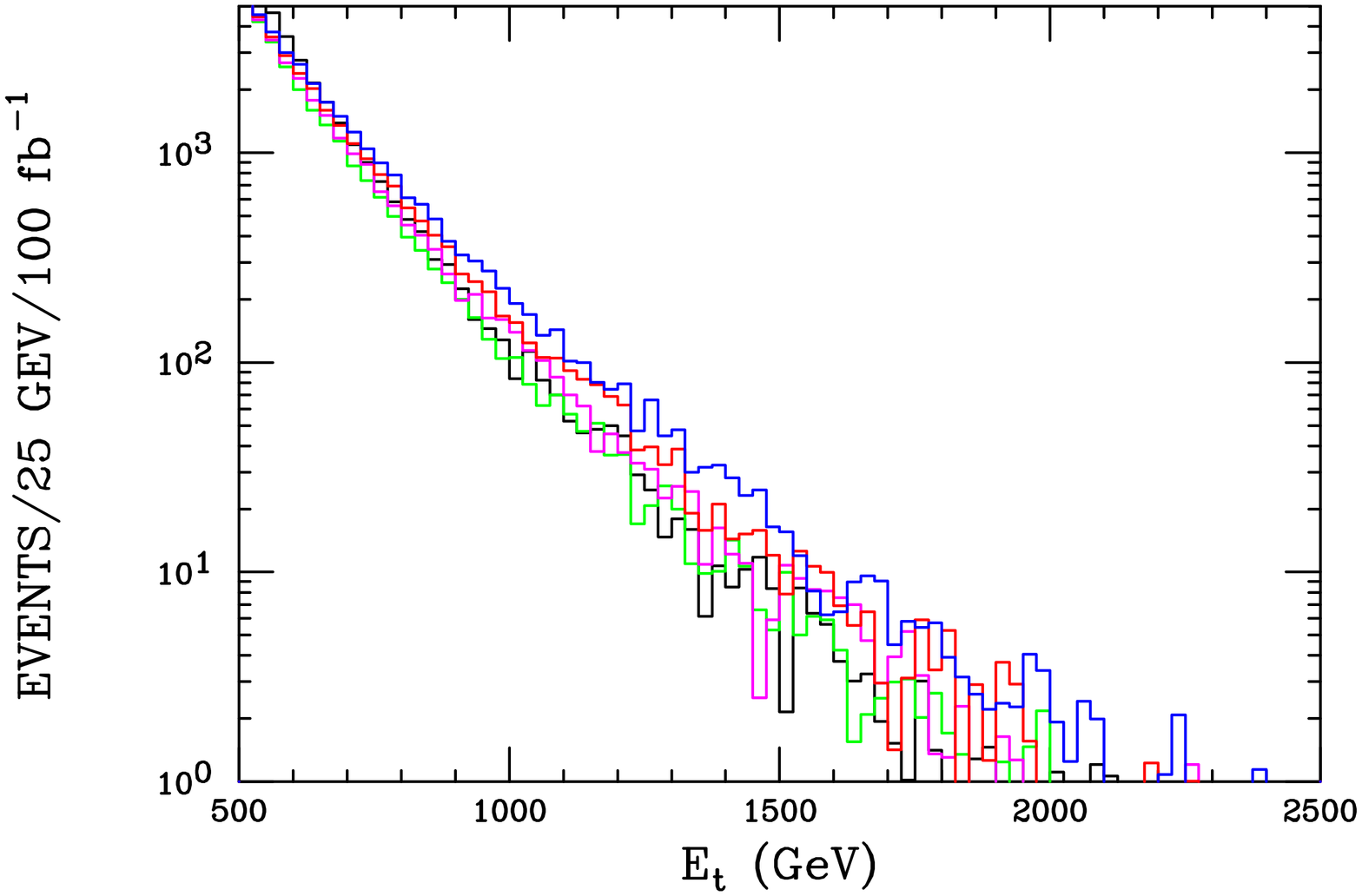}}
\vspace*{0.1cm}
\centerline{
\includegraphics[width=7.5cm,angle=90]{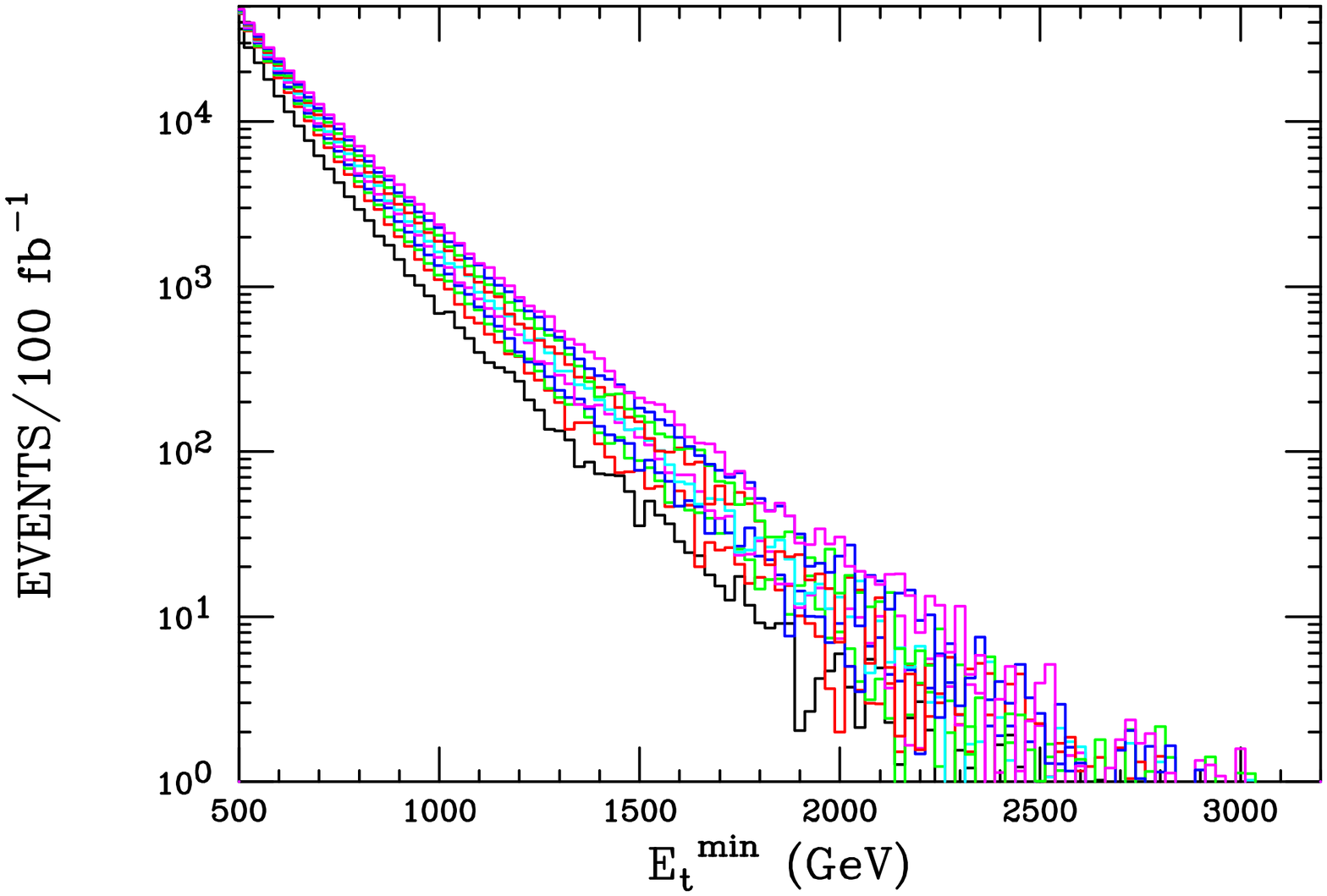}}
\vspace*{0.1cm}
\caption{Same as Fig.1, but now for $r=1$ scalar unparticles with, from bottom to top, $d=1.1$ to $d=1.9$ in steps of 0.1(0.2) in the lower(upper) panel.}
\label{fig4}
\end{figure}

Figs.~\ref{fig2}, ~\ref{fig3} and ~\ref{fig4} show the results analogous to Fig.~\ref{fig1} for the vector unparticle as well as the scalar unparticle 
cases with $r=0,1$, respectively. Several things are immediately obvious from these figures. First, as expected, the overall shape of the unparticle $E_T$ and  
$E_T^{cut}$ spectra becomes stiffer as the value of $d$ is increased. This is not too surprising as the leading subprocess cross sections are observed to scale 
as $\sim (\hat s/\Lambda^2)^{d-1}$ relative to the $Z+jet$ SM background. Secondly, the distributions we obtain for both the scalar and vector particles are, for 
the same value of $d$, essentially of the same shape. Again, this is not surprising as these two case differ only in the tensor structure of the unparticle 
interaction. The $r=0$ and $r=1$ scalar interactions are slightly different due to the additional unparticle coupling to the gluon. While this operator is 
suppressed by an additional power of $\hat s/\Lambda^2$, the initial state $gg$ luminosity at relatively low $E_T$ is very large and somewhat offsets this 
suppression. All in all, the predictions for the monojet spectra of these representative unparticle models are found to lie in a somewhat narrow band which 
is observed to be harder at high $E_T$ and $E_T^{cut}$ than is the SM background.

\begin{figure}[htbp]
\centerline{
\includegraphics[width=7.5cm,angle=90]{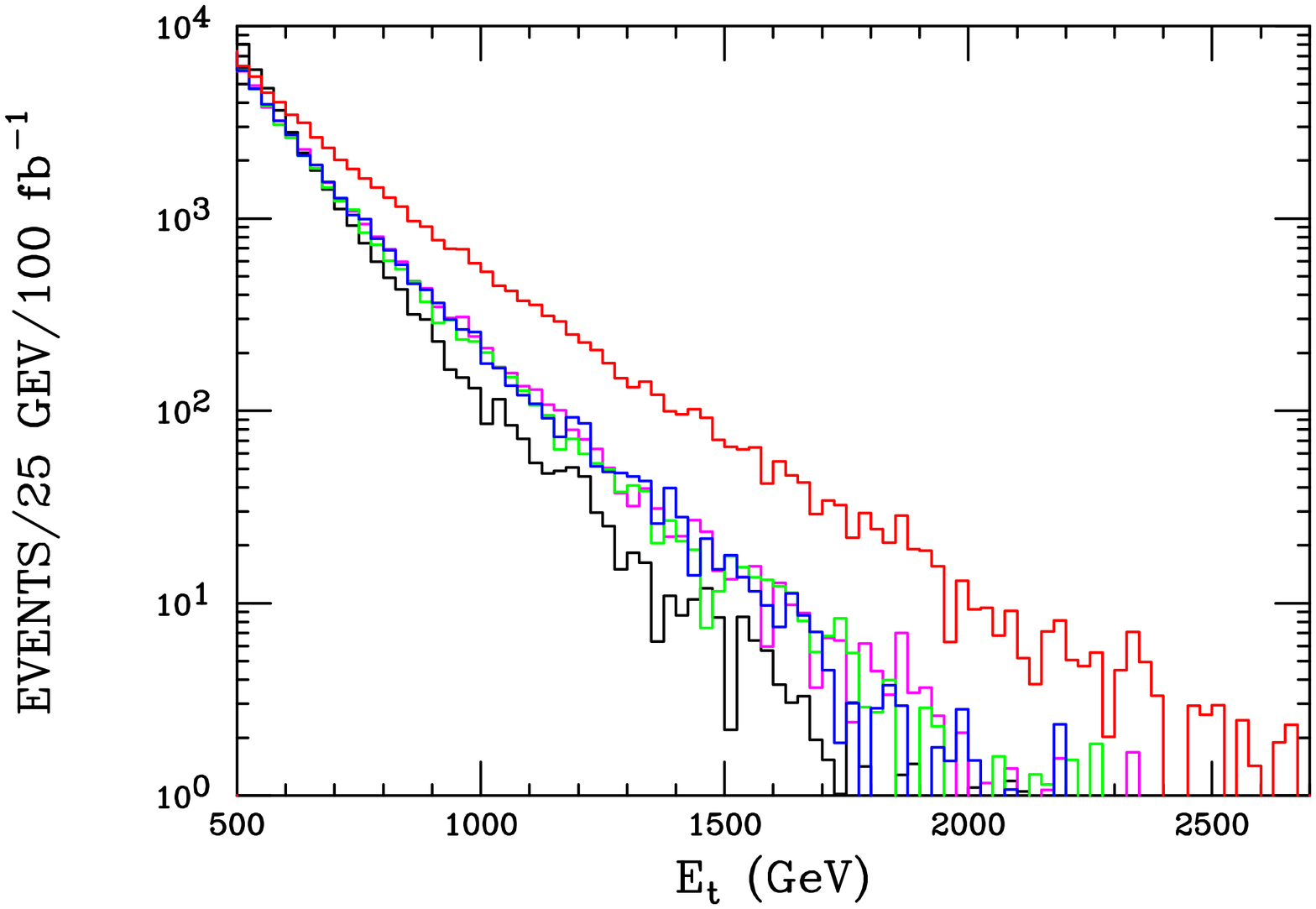}}
\vspace*{0.1cm}
\centerline{
\includegraphics[width=7.5cm,angle=90]{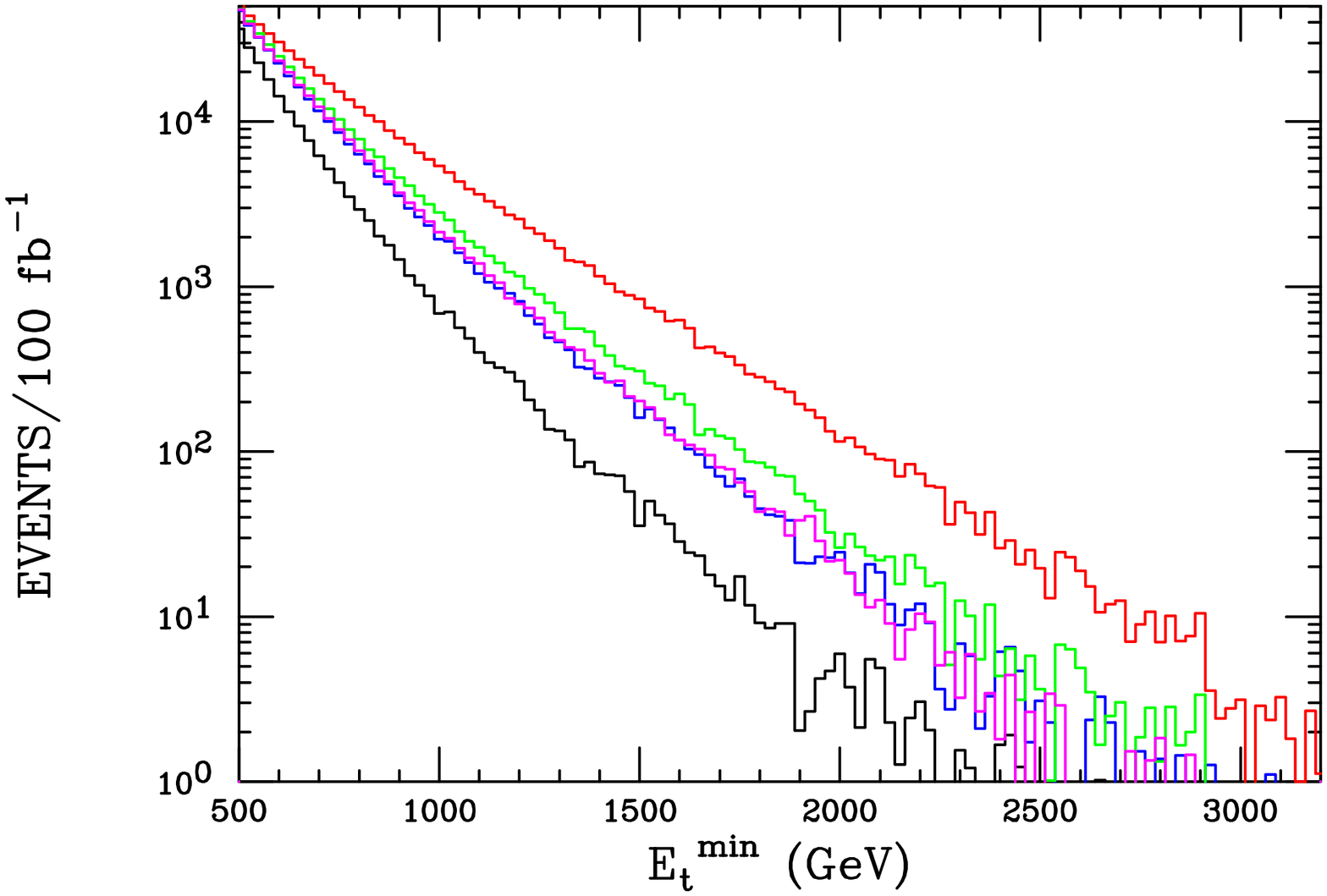}}
\vspace*{0.1cm}
\caption{Sample comparison of the predictions for the monojet $E_T$ and $E_T^{cut}$ distributions in the ADD model with $M_D=4$ TeV and $\delta=2$ (upper red 
histogram) with the case of vector Blue histogram) and scalar unparticles (with either $r=0,1$ in the green and magenta histograms, respectively) assuming 
$d=1.8$, corresponding to the. Also shown is the SM background (black histogram).}
\label{fig5}
\end{figure}

Up to now we have found that both the ADD predictions for the monojet $E_T$ and $E_T^{cut}$ spectra and those from unparticles separately 
lie in rather restricted ranges. 
How do these ranges compare? Since the hardest $E_T$ spectrum in the unparticle model is obtained for large $d$ and the softest spectrum for ADD occurs when 
$\delta=2$ it is most instructive to compare these specific two cases. Any other pair of spectra will clearly lie further apart in $E_T$ space and so will 
be more easily distinguished. Fig.~\ref{fig5} shows a representative set of comparisons of these two cases taking the above observations into account. Here 
we see that the three sample unparticle model predictions are relatively clustered together and lie in between those for the ADD scenario and the SM 
background. While the unparticle predictions themselves are difficult to distinguish it is clear that they are all easily differentiable from those of the ADD 
model. It is clear that a very large amount of jet energy smearing would be necessary to make these two predicted $E_T$ regions appear to overlap to any extent.
These results demonstrate that with enough statistics the shapes of the excess monojet $E_T$ distributions arising from these two classes of models can be 
distinguished at the LHC provided that they are visible above the SM background. We would, however, hope that a more detailed study by ATLAS/CMS using a full 
detector simulation will be performed to verify these results.

\section{Discussion and Conclusions}

In this paper we have explored the capability of the high luminosity LHC to differentiate two sets of new physics models that can lead to visible missing 
$E_T$/monojet signatures. The relevant tools for model discrimination are the shapes of the resulting monojet $E_T$ and $E_T^{cut}$ distributions themselves. 
First, we demonstrated that ($i$) the predictions of the ADD for the monojet $E_T/E_T^{cut}$ distributions form a very narrow band for $2 \leq \delta \leq 7$ 
and, ($ii$) Similarly, the corresponding 
predictions in the case of vector or scalar unparticles also so lie in a different but not so narrow band. Second, we showed that these two bands are seen not to 
overlap when integrated luminosities of order $\sim 100 fb^{-1}$ are available implying that these two classes of models which induce the monojet signal can be 
distinguished at the LHC.

%
%%%%%%%%%%%%%%%%%%--- References
%%%%%%%%%%%%%%%%%%%%%%%%%%%%%%%%%%%%%%%%%%%%%%%%%%%%%%%
\def\MPL #1 #2 #3 {Mod. Phys. Lett. {\bf#1},\ #2 (#3)}
\def\NPB #1 #2 #3 {Nucl. Phys. {\bf#1},\ #2 (#3)}
\def\PLB #1 #2 #3 {Phys. Lett. {\bf#1},\ #2 (#3)}
\def\PR #1 #2 #3 {Phys. Rep. {\bf#1},\ #2 (#3)}
\def\PRD #1 #2 #3 {Phys. Rev. {\bf#1},\ #2 (#3)}
\def\PRL #1 #2 #3 {Phys. Rev. Lett. {\bf#1},\ #2 (#3)}
\def\RMP #1 #2 #3 {Rev. Mod. Phys. {\bf#1},\ #2 (#3)}
\def\NIM #1 #2 #3 {Nuc. Inst. Meth. {\bf#1},\ #2 (#3)}
\def\ZPC #1 #2 #3 {Z. Phys. {\bf#1},\ #2 (#3)}
\def\EJPC #1 #2 #3 {E. Phys. J. {\bf#1},\ #2 (#3)}
\def\IJMP #1 #2 #3 {Int. J. Mod. Phys. {\bf#1},\ #2 (#3)}
\def\JHEP #1 #2 #3 {J. High En. Phys. {\bf#1},\ #2 (#3)}

\end{document}